**Extraction of crystal-field parameters for lanthanide ions from quantum-chemical calculations**

L. Hu[1,2]   M. F. Reid[*2]   C. K. Duan[3]   S. Xia[1]   M. Yin[1]

[1]Department of Physics, University of Science and Technology of China, Hefei 230026, China

[2]Department of Physics and Astronomy and MacDiarmid Institute for Advanced Materials and Nanotechnology, University of Canterbury, Christchurch 8140, New Zealand

[3]Institute of Modern Physics, Chongqing University of Post and Telecommunications, Chongqing 400065, China

PACS: 71.70.Ch, 71.15.–m, 71.23.An, 31.15.E-

Keywords: Crystal field; Lanthanides; Rare Earths; Energy levels

**Abstract**

A simple method for constructing effective Hamiltonians for the $4f^N$ and $4f^{N-1}5d$ energy levels of lanthanide ions in crystals from quantum-chemical calculations is presented. The method is demonstrated by deriving crystal-field and spin-orbit parameters for $Ce^{3+}$ ions doped in $LiYF_4$, $Cs_2NaYCl_6$, $CaF_2$, $KY_3F_{10}$ and YAG host crystals from quantum chemical calculations based on the DV-Xα method. Good agreement between calculated and fitted values of the crystal-field parameters is obtained. The method can be used to calculate parameters even for low-symmetry sites where there are more parameters than energy levels.

---

Corresponding authors:
* Email: mike.reid@canterbury.ac.nz (MFR);

## 1. Introduction

The $4f^N$ and $4f^{N-1}5d$ energy levels of trivalent lanthanide ions doped in crystals are important information for optical materials. Since the 1960s the $4f^N$ levels have been analysed by a parametric 'crystal-field' model.[1-5] More recently an extension of this model has been applied to the $4f^{N-1}5d$ levels.[6-10]

The parameters in the parametric 'crystal-field' model give valuable insight into the interactions of the 4f and 5d electrons with each other, with other electrons in the lanthanide ion, and with the crystalline environment.[5, 11] Furthermore, a key feature that makes parametric 'crystal field' calculations useful is that the parameters are found to vary predictably across the lanthanide series.[10] This means that once parameters have been determined for one or two ions it is relatively easy to extrapolate to other ions. These parametric analyses have not only provided a useful summary of the interactions mentioned above, but have also played a key role in the design of technological materials.[12]

The 'atomic' parameters in the 'crystal-field' model may be calculated by atomic many-body techniques[13] and spectroscopy of the $4f^N$ configuration has been an important test case for such calculations.[14] The crystal-field parameters may also be calculated, and it has been understood since the 1960s that the "crystal-field" is not just an electrostatic effect, but is the result of the complex interplay of quantum-mechanical effects.[15,16]

Reasonably accurate *ab initio* calculations of the lanthanide energy levels have become common in recent years.[17-22] Unfortunately, though quite good agreement can be obtained between *ab initio* calculations and experimental energy levels, most of these calculations do not derive crystal-field parameters. In some cases, particularly in high symmetries such as $O_h$, it is possible to determine the parameters from *ab initio* calculations by fitting the parameters to the calculated energy levels. However, this is not always possible in low symmetry, especially for $Ce^{3+}$ occupying a site of symmetry lower than $O_h$, $T_d$, or $D_6$, where there are an equal or greater number of free crystal-field and spin-orbit interaction parameters than energy level splittings. Parameters can provide a better test of the calculations because they can be compared with experimental parameters from ions other than the ion used in the calculation. So, for example, parameters calculated for $Ce^{3+}$ may be compared with experimental data for $Pr^{3+}$ and $Nd^{3+}$, as we will discuss below.

In this paper, we demonstrate how *ab initio* calculations may be used to determine parameters of a phenomenological effective Hamiltonian[23-25] such as the $4f^N$ and $4f^{N-1}5d$ crystal-field Hamiltonian. In this method, eigenvalues and eigenvectors from the *ab initio* calculations are used to construct the effective Hamiltonian. Preliminary results have been reported previously by Reid *et al.*[26,27] Here we present a detailed study of $Ce^{3+}$ in various host crystals with a variety of site symmetries. The calculated energy levels are compared with experimental measurements, and other quantum chemical calculations. We compare the crystal-field parameters with fitted parameters for $Ce^{3+}$ and/or other ions in the same crystal, such as $Pr^{3+}$ and $Nd^{3+}$, where the larger number of observable energy levels makes the fitted crystal-field parameters more reliable. Though we intend these calculations as a demonstration of principle, since the *ab initio* method we have used is relatively simple, the agreement between the calculations and experiment is generally good.

**2. Parametric crystal-field model**

The parametric crystal-field model of the $4f^N$ and $4f^{N-1}5d$ energy levels of trivalent and divalent lanthanide ions consists of an effective Hamiltonian operator $H_{\text{eff}}$, which acts on the multi-electron basis constructed from the single-electron orbitals of the ion under a central-field approximation. In principle, $H_{\text{eff}}$ can be as precise as the full $H$ in the sense of obtaining exact energies and projected wave functions in the model space, as long as all the effects are taken as effective interactions in the model space interactions. In practice, however, approximations are usually adopted in constructing $H_{\text{eff}}$.

The model Hamiltonian for the $4f^N$ configuration of a lanthanide ion in a crystal is commonly written as:[4]

$$H = H_{\text{free-ion}} + H_{\text{CF}}, \qquad (1)$$

where $H_{\text{free-ion}}$ includes the spherically-symmetric interactions present in the free lanthanide ion, and $H_{\text{CF}}$ describes the additional non-spherical symmetric interactions due to interactions of the lanthanide ion with the environment. The most common choices for $H_{\text{free-ion}}$ and $H_{\text{CF}}$ are:

$$H_{\text{free-ion}} = E_{\text{AVG}} + \sum_{k=2,4,6} F^k f_k + \zeta_{4f} A_{\text{SO}} + \alpha L(L+1) + \beta G(G_2)$$
$$+ \gamma G(R_7) + \sum_{i=2,3,4,6,7,8} T^i t_i + \sum_{k=0,2,4} M^k m_k + \sum_{k=2,4,6} P^k p_k \qquad (2)$$

$$H_{\text{CF}} = \sum_{k,q,i} B_q^k C_q^k(i) \qquad (3)$$

Here $E_{\text{AVG}}$ is a parameter that shifts the energy of the whole $4f^N$ configuration. $F^k$ and $f_k$ are electron repulsion parameters and operators. $\zeta_{4f}$ and $A_{\text{SO}}$ are the spin-orbit coupling constant and angular part of the spin-orbit interaction. $\alpha$, $\beta$ and $\gamma$ are two-particle configuration interaction parameters which are also known as Trees' parameters. $L$ is the total orbital angular momentum. $G(G_2)$ and $G(R_7)$ are Casimir operators for groups $G_2$ and $R_7$. The $T^i$ and $t_i$ are three-particle parameters and operators. The $M^k$ are Marvin integrals and $m_k$ the associated operators. The $P^k$ is electrostatic correlated spin-orbit interaction parameters and $p_k$ are the operators associated with $P^k$.

The $4f^N$ model Hamiltonian may be extended to the $4f^{N-1}5d$ configuration by including more interactions,[6,7] i.e.

$$H_{f^{N-1}d} = E_{\text{AVG}} + \sum_{k=2,4,6} F^k(\text{ff}) f_k(\text{ff}) + \zeta_{4f} A_{\text{SO}}(\text{ff}) + \alpha L(L+1) + \beta G(G_2) + \gamma G(R_7)$$
$$+ \sum_{i=2-4,6-8} T^i t_i + \sum_{k=0,2,4} M^k m_k + \sum_{k=2,4,6} P^k p_k + \sum_{k,q} B_q^k(\text{ff}) C_q^k(\text{ff}) + \Delta_E \delta_E(\text{fd}) +$$
$$\sum_{k=2,4} F^k(\text{fd}) f_k(\text{fd}) + \sum_{j=1,3,5} G^j(\text{fd}) g_j(\text{fd}) + \zeta_{5d} A_{\text{SO}}(\text{dd}) + \sum_{k,q} B_q^k(\text{dd}) C_q^k(\text{dd}). \qquad (4)$$

The parameters and operators have similar meanings to those in Eq. (2-3). The term $\Delta_E \delta_E(\text{fd})$

represents the difference between the average energy of the $4f^{N-1}5d$ configuration and the $4f^N$ configuration. The operator $\delta_E(fd)$ is diagonal, with unit matrix elements for $4f^{N-1}5d$ and zero matrix elements for $4f^N$. The $F^k(fd)$ and $G^j(fd)$ are direct and exchange Slater parameters for the Coulomb interaction between the 4f and 5d electrons. The $\zeta_{5d}$ parameter is associated with the spin-orbit interaction of the 5d electron. The $B_q^k(dd)$ is for the 5d electron affected by crystal-field interaction. More details are given in Refs. [6,7].

As explained in the Introduction, the 'crystal-field' interactions are not simply a result of electrostatic interactions between the ligands and the lanthanide ion, but arise from complex quantum-mechanical interactions between the 4f and 5d electrons and other electrons in the host crystal. Fitted crystal-field parameter values will automatically absorb contributions from all of such interactions.[5, 11, 15, 16]

All calculations in this paper use $Ce^{3+}$ as the lanthanide ion. In this case there is only one valence electron and the 4f and 5d configurations that we consider only involve one-electron operators (i.e., crystal-field and spin-orbit interactions). Consequently, the Hamiltonian simplifies to:

$$H = \zeta(ff)A_{SO} + \sum_{k,q} B_q^k(ff)C_q^k(ff) + \Delta_E(fd)\delta_E(fd) + \zeta(dd)A_{SO} + \sum_{k,q} B_q^k(dd)C_q^k(dd). \quad (5)$$

## 3. DV-Xα calculations

To demonstrate our method we make use of the DV-Xα computer program. This program was originally developed for quantum chemical calculations of electronic and structural properties on molecular systems by D. E. Ellis and co-workers,[28] and was further developed by Adachi and co-workers.[29] The program employs the discrete variational method to a cluster isolated or embedded in microcrystal. α is the parameter for exchange-correlation potential and fixed to be 0.7 in this work. The DV-Xα method gains numerical efficiency by replacing the calculation of integrals by summation of data over sampling points that are appropriate for the charge distribution of the system in question. It has been used not only to calculate the single-electron states of lanthanides,[30] but also to obtain molecular orbitals used in the calculation of eigenvalues and eigenvectors of many-electron states.[17] In this work we use a relativistic version, which therefore automatically includes the spin-orbit interaction.

In the relativistic DV-Xα method, the one-electron states are obtained by solving the one-electron Dirac equation:

$$H\phi_k(\mathbf{r}) = E_k \phi_k(\mathbf{r}) \tag{6}$$

where $r$ is the position of the electron, and $\Phi_k(r)$ and $E_k$ are the $k$th molecular orbital and its energy. $H$ represents the one-electron Dirac Hamiltonian, Eq. (2) of Chapter 1 in Ref. [17]. The $k$th molecular orbital $\Phi_k(r)$ is expressed as a linear combination of atomic orbitals (LCAO), i.e.

$$\phi_k(\mathbf{r}) = \sum_i C_{ik} \varphi_i(\mathbf{r}), \tag{7}$$

where $\varphi_i(r)$ is the $i$th atomic orbital.

## 4. Crystal-field parameters from DV-Xα calculations

A method for obtaining effective Hamiltonian $H_{\text{eff}}$ in a model space from a quantum-chemical *ab initio* / first principle calculations was given by Reid *et al.*[26] The advantage of this method is that it makes uses not only of the energies but also of the model-space projection of the eigenvectors. This allows us to determine the parameters even in cases where there are more parameters (spin-orbit, crystal-field, *etc.*) in the parametric Hamiltonian than the number of energy levels.

To calculate the crystal-field parameters for $Ce^{3+}$ in crystals, all 14 $^2F_{JM}$ (or 10 $^2D_{JM}$) basis functions are used to form the 'model space'. In the DV-Xα calculation the output wave functions $\varphi(r)$ in (6) are given as linear combinations of those bases and other orbitals, such as the 6s and 5p orbitals of $Ce^{3+}$ and the orbitals of the ligand atoms.

To construct the effective Hamiltonian for the levels that can be considered as '4f' (or '5d') levels of $Ce^{3+}$, the 14 (or 10) energy levels with the largest 4f (or 5d) components are chosen. We denote the diagonal matrix constructed with diagonal elements the 14 (or 10) energies as $\mathbf{E}_p$ and the matrix for the eigenvectors projected into the 'model' space as $\mathbf{V}_p$. The method presented in Ref. [26] can then be used to construct the effective Hamiltonian. This can then be used to calculate crystal-field parameters. A brief discussion of the method is given below. Further details may be found in Ref. [26].

The projected eigenvectors are, in general, not orthonormal but can usually be expected to be non-singular. An orthonormal matrix $\mathbf{V}_k$ can be constructed from $\mathbf{V}_p$ as:

$$\mathbf{V}_k = ((\mathbf{V}_p \mathbf{V}_p^\dagger)^{-1})^{1/2} \mathbf{V}_p . \tag{8}$$

Then an Hermitian effective Hamiltonian can be constructed as:

$$\mathbf{H}_{\text{eff}} = \mathbf{V}_k \mathbf{E}_p \mathbf{V}_k^{-1} . \tag{9}$$

It can be seen immediately that the eigenvalues of $\mathbf{H}_{\text{eff}}$ are the diagonal elements of $\mathbf{E}_p$ and the eigenvectors of $\mathbf{H}_{\text{eff}}$ are $\mathbf{V}_k$. Following Ref. [26], $\mathbf{H}_{\text{eff}}$ can be expanded in terms of a complete set of operators $\mathbf{T}_\alpha$ describing the effective interactions in the model space as:

$$\mathbf{H}_{\text{eff}} = \sum_\alpha P_\alpha \mathbf{T}_\alpha . \tag{10}$$

Here $P_\alpha$ are the parameters to describe the strength of the various effective interactions:

$$P_\alpha = \sum_\beta (\mathbf{A}^{-1})_{\alpha\beta} \operatorname{tr}(\mathbf{T}_\beta^\dagger \mathbf{H}_{\text{eff}}) , \tag{11}$$

where $\mathbf{A}$ is a matrix with elements:

$$A_{\alpha\beta} = \operatorname{tr}(\mathbf{T}_\alpha^\dagger \mathbf{T}_\beta) . \tag{12}$$

For some site symmetries not all crystal-field parameters can be chosen to be real. The crystal-field parameters are modified by rotations of the axis system, as described in Ref. [4]. All $B_0^k$ are real due to the hermiticity and time-reversal symmetries of the crystal-field interaction.[31] Rotating about the $z$ axis by angle $\varphi$ gives the following change in phase for the parameters:

$$B_q^k = B_q^k e^{-iq\varphi} . \tag{13}$$

Hence at least one $B_q^k$ ($q\neq 0$) may be chosen to be real, and it has been show that is always possible to perform rotations to make all $k=2$ parameters real.[32]

In practice, most calculations based on parameter fitting have been carried out by assuming that all parameters are real. Where our calculated crystal-field parameters are complex, we take the sign from the real part of our calculated parameters. In this paper this is only an issue for LiYF$_4$. In that case, for a suitable choice of axes, the only parameter with an imaginary part is $B_4^6$, and the imaginary part is calculated to be small. In several cases we use Eq. (13) to change the phases of parameters to match them to the choice of axes made in our calculation.

In the following calculations we are primarily interested in crystal-field splitting. Therefore the calculated 5d energies are adjusted by shifting all levels by a constant amount so that the average of 5d energies matches the experimental average.

## 5. Examples

*5.1 LiYF$_4$:Ce$^{3+}$*

We use LiYF$_4$:Ce$^{3+}$ to illustrate our method. DV-Xα calculations for a number of lanthanide ions in LiYF$_4$ have been reported in Ref. [33], which we can compare our calculations to. Another reason for choosing this system is that Ce$^{3+}$ occupies a low-symmetry $S_4$ site in LiYF$_4$. Consequently, the number of crystal-field and spin-orbit parameters is larger than the number of energy levels and so it is not possible to determine all of the parameters by a parametric fitting of only the experimental or calculated *energies*. However, the method described in Sec. 4 can be used to determine all of the parameters.

In order to investigate the effect of different cluster sizes, calculations were carried out for three different clusters (CeF$_8$)$^{5-}$, (CeLi$_4$F$_{12}$)$^{5-}$ and (CeY$_4$Li$_8$F$_{12}$)$^{11+}$. All the clusters were embedded in a microcrystal containing about 1300 atoms. The coordinates of atoms were taken from Ref. [34], and differ slightly from those used in Ref. [27].

Table 1 lists experimental energies, energies calculated by Watanabe *et al.* in Chapter 5 of Ref. [17] and energies from our calculations. It can be seen that the calculated 5d splitting for the small $(CeF_8)^{5-}$ cluster is quite close to those of Refs. [17, 22], which were obtained with the same method but a slightly different choice of basis functions. In the calculation by S. Watanabe all orbitals from 1s to 6p for cerium and from 1s to 2p for fluorine were allowed to vary, while in our calculation only 4f, 5p, 5d and 6s for cerium and 2s, 2p for fluorine were allowed to vary, and inner orbitals were frozen in order to speed up the calculation. This may explain why our calculations tend to underestimate the average energy of the 5d configuration, whereas the calculations of Refs. [17, 22] tend to overestimate it. Table 2 lists the available experimental crystal-field and spin-orbit parameters for $Ce^{3+}$, $Pr^{3+}$, and $Nd^{3+}$ in $LiYF_4$ for the $4f^N$ and $4f^{N-1}5d$ configurations. In the case of $Ce^{3+}$ the 4f parameters are extrapolated from those for $Pr^{3+}$ and $Nd^{3+}$ ions due to the lack of experimental data. For the small cluster the calculated crystal-field and spin-orbit parameters are quite consistent with the experimental parameters. However, for larger clusters the parameters are no longer consistent with those obtained by fitting experimental data. We conclude that the calculations for the large clusters are physically unrealistic. This may be illustrated by analysing the states in terms of atomic orbital percentages. These are plotted in Figure 1 for the largest cluster adopted in our calculations. In this case the lowest 5d state is 73% 5d components, with most other components from orbitals of $Y^{3+}$, but all the other states contain less than 50% 5d components. While experimental data such as line-widths and photoconductivity[35] suggest that the higher 5d states should mix with the conduction band (being built primarily from $Y^{3+}$ orbitals) the mixing in our calculation is not physically realistic.

This delocalization has been noted by Pascual *et al.*[19] In Figure 1 of Ref. [19] it is demonstrated that a simple Madelung embedding (such as the one used here) can lead to an unphysical delocalization of the high-energy electrons. A more robust quantum-mechanical embedding such as the one implemented in Ref. [19] would be required to treat these states more accurately.

Since our calculations for large clusters give unrealistic results, for other systems we only present results for clusters that explicitly consider only $Ce^{3+}$ and it's nearest neighbours. In fact, almost all *ab initio* calculations on lanthanide ions in crystals have used such small clusters.

*5.2 $Cs_2NaYCl_6$*

The space group of $Cs_2NaYCl_6$ is Fm3m (SG number 225).[36] $Ce^{3+}$ substitutes for $Y^{3+}$ on sites of octahedral symmetry. For 4f and 5d states, only two and one crystal-field parameters are required respectively. The crystal-field splitting of 4f and 5d states alone can be used to fully determine the crystal-field parameters, making the method summarized in Sec. 4 unnecessary. Nevertheless, it is useful to test DV-Xα calculation in such high symmetry systems. Calculated and experimental energy levels and parameters are listed in Table 3 and Table 4. It can be seen that our calculations are consistent with experimental values and with the calculation of Ref. [37].

Note that for octahedral symmetry the ratios of certain crystal-filed parameters are fixed, $B_4^4/B_0^4 = \sqrt{5/14}$ and $B_4^6/B_0^6 = -\sqrt{7/2}$ ,[38] so in this case, and for $CaF_2$, we only give $B_0^4$ and $B_0^6$ parameters.

*5.3 $CaF_2$*

The space group of $CaF_2$ is Fm3m (SG number 225).[39] We consider the case that $Ce^{3+}$ substitutes for $Ca^{2+}$ and maintains the cubic site symmetry (i.e., the charge compensation ion or vacancy is far away). [40] Table 5 presents the calculated 4f and 5d energy levels and Table 6 presents calculated and experimental parameters. The calculated splitting of 5d energy levels is smaller than the experimental one. This may be because the substitution of divalent $Ca^{2+}$ with trivalent $Ce^{3+}$ will result in a reduction of the Ce-F distance.

*5.4 KY$_3$F$_{10}$*

The space group of KY$_3$F$_{10}$ is Fm3m (SG number 225).[41] Ce$^{3+}$ substitutes Y$^{3+}$ in a site of $C_{4v}$ symmetry. In Table 7 we compare our calculated energy levels with those from Chapter 5 of Ref. [17]. Since there are no experimental parameters available for Ce$^{3+}$ in this host, parameters obtained from fitting experimental data for Pr$^{3+}$ and Nd$^{3+}$ are shown in Table 8, together with those calculated here for Ce$^{3+}$. Considering that the crystal-field parameters for Ce$^{3+}$ are expected to be slightly larger than those for Pr$^{3+}$ and Nd$^{3+}$ ions it can be seen that the agreement between calculated an experimental parameters is quite satisfactory.

*5.5 YAG*

The space group of YAG is Ia3d (SG number 230).[42] Ce$^{3+}$ replaces Y$^{3+}$ site of $D_2$ symmetry. Table 9 lists the results of energies and Table 10 lists the parameters. There are six equivalent sets of parameters for $D_2$ symmetry due to six different choices of coordinate system compatible with $D_2$ symmetry. Detailed discussion of this point is given by Morrison and Leavitt.[43] In general, it is not possible to determine the correspondence of the parameter set with coordinate system from parametric fitting, unless there are other information, such as EPR data, available.

In our calculation, once a coordinate system was chosen and the ions positions were given, a unique set of energies and eigenvectors were obtained and hence a unique set of parameters would be obtained. Comparing the six equivalent experimental parameter sets we find that the set 3 of Ref. [43] corresponds to our choice of coordinate system for the calculations. However, we have adjusted signs to match the calculations by rotating the system, as explained in section 4. The agreement with experiment is generally satisfactory for the 5d and 4f parameters, except for some 4f crystal-field parameters with relative small values.

**6. Conclusions**

We have demonstrated that crystal-field parameters for $4f^N$ and $4f^{N-1}5d$ configurations of lanthanide ions in crystals can be obtained systematically from *ab initio* calculations. The method we have used involves the construction of the effective Hamiltonian making use of both eigenvalues and projected eigenvectors from the ab-initio calculation. The use of the eigenvectors means that the method is applicable in cases of low symmetry, where there are more parameters than energy levels.

Our calculation uses a relatively simple DV-Xα method, and a simple Madelung embedding. However, the calculated results are quite consistent with the experimental data. They are accurate enough that it should be possible to use the method for the calculations of low-symmetry $Ce^{3+}$ systems for which crystal-field parameters can not be determined from fitting experimental data. This is an important application, as there is considerable interest in phosphors that make use of the 5d → 4f transitions.[44, 45]

Since crystal-field parameters are transferable to other ions in the lanthanide series, another application of our approach is to derive parameters from calculations from $Ce^{3+}$, for which *ab initio* calculations are manageable than for a system with many 4f electrons, and then scale those parameters for crystal-field calculations for other ions.

Applying our method of extracting parameters to more sophisticated calculations or higher accuracy gives the possibility of investigating the physics of the crystal-field interactions in more detail. For example, it would be interesting to examine the dependence of calculated crystal-field parameters across the lanthanide series and compare the trends with experimental data.[46] It would also be interesting to investigate the dependence of the parameters on bond distances and angles, as has been discussed in Refs. [11, 47]. Such analyses would provide useful tests of both the *ab initio* calculations and the various empirical models that are used to analyse experimental data and predict other spectroscopic properties.[4,5,11]

Our method can also be expanded to constructing effective operators for other properties of the system. Of particular interest are the electric dipole transition intensities within the $4f^N$ configuration, which is commonly addressed via a parametric 'Judd-Ofelt' model.[48,49] Few attempts of true *ab initio* evaluation of these transition intensities have been attempted,[50,51] with most theoretical work using simple point-charge and induced-dipole ("dynamic coupling") mechanisms.[4, 6, 52] The Judd-Ofelt parameters may be determined from effective dipole-moment operators. Such calculations will be the subject of a future study.

**Acknowledgements**

L. Hu acknowledges financial support from China Scholarship Council that enabled his visit to New Zealand and the National Natural Science Foundation of China under grant Nos. 11011120083, 10904139 and 11074315.

S. Xia acknowledges supports of the national science foundation of China under grant Nos. 10874173 and 11074245.

M. Yin acknowledges financial support of the Knowledge Innovation Project of The Chinese Academy of Sciences under grant No. KJCX2-YW-M11, and the Special Foundation for Talents of Anhui Province, China under grant No. 2007Z021.

**References**

[1] B. G. Wybourne, *Spectroscopic Properties of Rare Earths* (Interscience, New York, 1965).

[2] G. H. Dieke, *Spectra and Energy Levels of Rare Earth Ions in Crystals* (Interscience, New York, 1968).

[3] W. T. Carnall, P. R. Fields, K. Rajnak, *J. Chem. Phys.* **49**, 4424 (1968).

[4] G. K. Liu, B. Jacquier (Eds.), *Spectroscopic Properties of Rare Earths in Optical Materials* (Springer, New York, 2005).

[5] D. J. Newman, B. Ng, *Crystal Field Handbook* (Cambridge University Press, Cambridge, 2000).

[6] G. W. Burdick, M. F. Reid, *Handbook on the physics and chemistry of rare earth*, vol. 37 (Eds: K. A. Gschneidner, Jr., J. –C. G. Bünzli, V. K. Pecharsky), chapter 232, North Holland, Amsterdam, 61 (2007).

[7] M. F. Reid, L. van Pieterson, R. T. Wegh, A. Meijerink, *Phys. Rev. B* **62**, 14744 (2000).

[8] M. Karbowiak, A. Urbanowicz, M. F. Reid, *Phys. Rev. B* **76**, 115125 (2007).

[9] C. K. Duan, P. A. Tanner, *J. Phys.: Condens. Matter* **20,** 215228 (2008).

[10] C. K. Duan, P.A. Tanner, *J. Phys. Chem. A* **114**, 6055 (2010).

[11] D. J. Newman, B. Ng, *Rep. Prog. Phys.*, **52**, 699 (1989).

[12] B. Henderson, R. H. Bartram, *Crystal-Field Engineering of Solid-State Laser Materials,* (Cambridge University Press, Cambridge, 2000).

[13] J. C. Morrison, K. Rajnak, *Phys. Rev. A* **4**, 536 (1971).

[14] E. Eliav, U. Kaldor, Y. Ishikawa, *Phys. Rev. A* **51**, 225 (1995).

[15] M. M. Ellis, D. J. Newman, *Phys. Lett.* **21**, 508 (1966).

[16] B. Ng, D. J. Newman, *J. Chem. Phys.* **87**, 7110 (1987).

[17] M. G. Brik, K. Ogasawara, *First-principles Calculations of Spectral Properties of Rare Earth and Transition Metal Ions in Crystals* (Transworld Research Network, Kerala, 2006).

[18] F. Ruiperez, A. Barandiarán, L. Seijo, *J. Chem. Phys.* **123**, 244703 (2005).

[19] J. L. Pascual, J. Schamps, Z. Barandiarán, L. Seijo, *Phys. Rev. B* **74**, 104105 (2006).

[20] G. Sánchez-Sanz, L. Seijo, Z. Barandiarán, *J. Chem. Phys.* **131**, 024505 (2009).

[21] J. Andriesson, E. van der Kolk, P. Dorenbos, *Phys. Rev. B* **76**, 075124 (2007).

[22] S. Watanabe, T. Ishii, K. Fujimaru, K. Ogasawara, *J. Solid State Chem.* **179**, 2438 (2006).

[23] V. Hurtubise, K. F. Freed, *Adv. Chem. Phys.* **83**, 465 (1993).

[24] A. R. Bryson, M. F. Reid, *J. Alloys Comp.* **275-277**, 284 (1998).

[25] M. J. Lee, M. F. Reid, M. D. Faucher, G. W. Burdick, *J. Alloys Comp.* **323-324**, 636 (2001).

[26] M. F. Reid, C. K. Duan, H. W. Zhou, *J. Alloys Comp.* **488**, 591 (2009).

[27] M. F. Reid, L. Hu, S. Frank, C. K. Duan, S. Xia, M. Yin. *Eur. J. Inorg. Chem.* **18**, 2649 (2010).

[28] A. Rosen, D. E. Ellis, *J. Chem. Phys.* **62**, 3039 (1975).

[29] H. Adachi, M. Tsukada, C. Satoko, *J. Phys. Soc. Japn.* **45**, 875 (1978).

[30] S. Nomura, T. Takizawa, S. Endo, M. Kai, *Phys. Stat. Sol. (c)* **8,** 2739 (2006).

[31] D. J. Newman, *Adv. Phys.*, **20**, 197 (1971).

[32] M. F. Reid, G. W. Burdick, H. J. Kooy, *J. Alloys Comp.* **207-208**, 78 (1994).

[33] K. Ogasawara, S. Watanabe, H. Toyoshima, T. Ishii, M. G. Brik, H. Ikeno, I. Tanaka, *J. Solid State Chem.* **178**, 412 (2005).

[34] A. V. Goryunov, A. I. Popov, N. M. Khaidukov, P. P. Fedorov, *Mater. Res. Bull.* **27**, 213 (1992).

[35] L. van Pieterson, M. F. Reid, R. T. Wegh, S. Soverna, A. Meijerink, *Phys. Rev. B* **65**, 045113 (2002).

[36] C. Reber, H. U. Guedel, G. Meyer, T. Schleid, C. A. Daul, *Inorg. Chem.* **28**, 3249 (1989).

[37] P. A. Tanner, C. S. K. Mak, N. M. Edelstein, K. M. Murdoch, G. Liu, J. Huang, L. Seijo, Z. Barandianrán, *J. Am. Chem. Soc.* **125**, 13225 (2003).

[38] C. Görller-Walrand, K. Binnemans, *Handbook on the Physics and Chemistry of Rare Earths*, vol. 23 (Eds.: K. A. Gschneidner Jr., L. Eyring), North-Holland, Amsterdam, 121, (1996).

[39] A. K. Cheetham, B. E. F. Fender, M. J. Cooper, *J. Phys. C* **4**, 3107 (1971).

[40] L. van Pieterson, R. T. Wegh, A. Meijerink, M. F. Reid, *J. Chem. Phys*. **115**, 9382 (2001).

[41] E. F. Bertaut, Y. Fur, S. Aleonard, *Z. Kristallogr.* **187**, 279 (1989).

[42] A. Nakatsuka, A. Yoshiasa, T. Yamanaka, *Acta Crystallogr. B* **55**, 266 (1999).

[43] C. A. Morrison, R. P. Leavitt, *Handbook on the Physics and Chemistry of Rare Earths*, vol. 5 (Eds.: K. A. Gschneidner, L. Eyring), North-Holland, Amsterdam, 461 (1996).

[44] J. Lü, Y. Huang, Y. Tao, H. J. Seo, J. Alloys Comp. 500, 134 (2010).

[45] Z. Tian, H. Liang, W. Chen, Q. Su, G. Zhang, G. Yang, *Opt. Express* **17**, 957 (2009).

[46] W. T. Carnall, G. L. Goodman, K. Rajnak, R. S. Rana, *J. Chem. Phys.* **90**, 3443(1989).

[47] B. Ng, D. J. Newman, *J. Phys. C: Solid State Phys.* 19: L585 (1986).

[48] B. R. Judd, *Phys. Rev.* **127**, 750 (1962).

[49] G. S. Ofelt, *J. Chem. Phys.* **37**, 511 (1962).

[50] M. F. Reid, B. Ng, *Mol. Phys.* **67,** 407 (1989).

[51] M. Kotzian, T. Fox, N. Rosch, *J. Phys. Chem.* **99**, 600 (1995).

[52] M. F. Reid, J. J. Dallara, F. S. Richardson, *J. Chem. Phys.* **79**, 5743 (1983).

[53] G. W. Burdick, F. S. Richardson, *J. Alloys Comp.* **275-277**, 379 (1998).

[54] J. P. R. Wells, M. Yamaga, T. P. J. Han, H. G. Gallagher, *J. Phys.: Condens. Matter* **12**, 5297 (2000).

[55] M. Mujaji, J. R. Wells, *J. Phys.: Condens. Matter* **21**, 255402 (2009).

[56] P. A. Tanner, L. Fu, L. Ning, B. M. Cheng, M. G. Brik, *J. Phys.: Conden. Matter* **19**, 216213 (2007).

[57] J. Gracia, L. Seijo, Z. Barandiarán, D. Curulla, H. Neimansverdriet, W. van Gennip, *J. Lumin.* **128**, 1248 (2008).

[58] C. A. Morrison, D. E. Wortman, N. Karayianis, *J. Phys. C: Solid State Phys.*, **9**, L191 (1976).

Table 1. 4f and 5d energy levels of $Ce^{3+}$ in $LiYF_4$. Units are $cm^{-1}$. The calculated energies for the 5d states have been shifted by a constant amount so that the average matches the experimental average of 45277 $cm^{-1}$. Before this correction the calculated averages were 49490 $cm^{-1}$, 30864 $cm^{-1}$ 48480 $cm^{-1}$ and 81176 $cm^{-1}$ respectively.

|  | Experiment[a] | Calculation[b] | $(CeF_8)^{5-}$ | $(CeLi_4F_{12})^{5-}$ | $(CeLi_8Y_4F_{12})^{11+}$ |
|---|---|---|---|---|---|
| 4f[1] |  | 0 | 0 | 0 | 0 |
|  |  | 129 | 368 | 184 | 556 |
|  |  | 492 | 414 | 466 | 1347 |
|  |  | 2807 | 2650 | 2642 | 3184 |
|  |  | 2896 | 2777 | 2657 | 3506 |
|  |  | 3041 | 3010 | 3091 | 4289 |
|  |  | 3646 | 3265 | 3094 | 4573 |
| 5d[1] | 33433 | 35873 | 33348 | 35395 | 31525 |
|  | 41101 | 41760 | 42179 | 41781 | 40807 |
|  | 48564 | 47568 | 48339 | 47076 | 49308 |
|  | 50499 | 48616 | 48957 | 47660 | 49641 |
|  | 52790 | 52568 | 53564 | 54475 | 55102 |

[a] Experiment energy levels from Ref. [40].

[b] DV-Xα calculation from Chapter 5 of Ref. [17].

Table 2. Crystal-field parameters and spin-orbit interaction parameters of $Ce^{3+}$, $Pr^{3+}$ and $Nd^{3+}$ in $LiYF_4$. Units are $cm^{-1}$.

| | Parameters | $Ce^{3+[a]}$ | $Pr^{3+[b]}$ | $Nd^{3+[c]}$ | $(CeF_8)^{5-}$ | $(CeLi_4F_{12})^{5-}$ | $(CeLi_8Y_4F_{12})^{11+}$ |
|---|---|---|---|---|---|---|---|
| 4f | $B_0^2(ff)$ | 481 | 489 | 409 | 618 | -819 | 2404 |
| | $B_0^4(ff)$ | -1150 | -1043 | -1135 | -538 | -435 | 1350 |
| | $B_4^4(ff)$ | -1228 | -1242 | -1216 | -966 | -485 | -874 |
| | $B_0^6(ff)$ | -89 | -42 | 27 | 143 | 466 | -345 |
| | $B_4^6(ff)$ | -1213 | -1213 | -1083 | -818 | -737 | -1071 |
| | $\zeta_{4f}$ | 615 | 731 | 871 | 752 | 752 | 910 |
| 5d | $B_0^2(dd)$ | 4673 | 7290 | | 4075 | -612 | 3339 |
| | $B_0^4(dd)$ | -18649 | -14900 | | -14296 | -11012 | -19136 |
| | $B_4^4(dd)$ | -23871 | -17743 | | -25162 | -23783 | -29128 |
| | $\zeta_{5d}$ | 1082 | 906 | | 841 | 773 | 365 |

[a] Extrapolated and fitted parameters from Ref. [40].

[b] Fitted parameters from Ref. [7]. Signs of $B_4^4(ff)$ and $B_4^6(ff)$ and $B_4^4(dd)$ have been changed to match the axis choice of our calculation by rotating π/4 about z axis.

[c] Fitted parameters from Ref. [38].

Table 3. 4f and 5d energy levels of $Ce^{3+}$ in $Cs_2NaYCl_6$ (unit: $cm^{-1}$). The calculated energies in this work for the 5d states have been shifted by a constant amount so that the average matches the experimental average of 36015 $cm^{-1}$. Before this correction the calculated average was 13806 $cm^{-1}$.

|  | Irrep | Degeneracy | Experiment[a] | Calculation[b] | Calculation[c] |
|---|---|---|---|---|---|
| 4f$^1$ | $\Gamma_{7u}$ | 2 | 0 | 0 | 0 |
|  | $\Gamma_{8u}$ | 4 | 597 | 831 | 598 |
|  | $\Gamma_{7u}$ | 2 | 2167 | 2318 | 2536 |
|  | $\Gamma_{8u}$ | 4 | 2691 | 3019 | 3053 |
|  | $\Gamma_{6u}$ | 2 | 3085 | 3376 | 3478 |
| 5d$^1$ | $\Gamma_{8g}$ | 4 | 28196 | 25510 | 29244 |
|  | $\Gamma_{7g}$ | 2 | 29435 | 26716 | 30312 |
|  | $\Gamma_{8g}$ | 4 | 47125 | 47263 | 45638 |

[a] Experiment energy levels from Ref. [37].

[b] *Ab initio* embedded cluster calculations from Ref. [37].

[c] DV-Xα calculated on $(CeCl_6)^{3-}$ embedded in $Cs_2NaYCl_6$ microcrystal in this work.

Table 4. Crystal-field parameters and spin-orbit interaction parameters of $Ce^{3+}$, $Pr^{3+}$, and $Nd^{3+}$ in $Cs_2NaYCl_6$. Units are $cm^{-1}$.

|  | Parameters | $Ce^{3+}$[a] | $Pr^{3+}$[b] | $Nd^{3+}$[c] | Calculation[d] |
|---|---|---|---|---|---|
| 4f | $B_0^4(ff)$ | 2208 | 2279 | 1966 | 2230 |
|    | $B_0^6(ff)$ | 250 | 293 | 258 | 310 |
|    | $\zeta_{4f}$ | 624 | 747 | 872 | 732 |
| 5d | $B_0^4(dd)$ | 38709 |  |  | 33530 |
|    | $\zeta_{5d}$ | 793 |  |  | 683 |

[a] Fitted parameters from Ref. [37].

[b] Fitted parameters from Ref. [53].

[c] Fitted parameters from Ref. [43].

[d] DV-Xα calculated on $(CeCl_6)^{3-}$ embedded in $Cs_2NaYCl_6$ microcrystal in this work.

Table 5. 4f and 5d energy levels of $Ce^{3+}$ in $CaF_2$ (unit: $cm^{-1}$). The calculated energies for the 5d states have been shifted by a constant amount so that the average matches the experimental average of 44929 $cm^{-1}$. Before this correction the calculated averages were 48119 $cm^{-1}$ and 54145 $cm^{-1}$ respectively.

|  | Irrep | Degeneracy | Experiment[a] | Calculation[b] | Calculation[c] |
|---|---|---|---|---|---|
|  | $\Gamma_{8u}$ | 4 | 0 | 0 | 0 |
|  | $\Gamma_{7u}$ | 2 | 430 | 468 | 420 |
| $4f^1$ | $\Gamma_{8u}$ | 4 | 2106 | 2750 | 2910 |
|  | $\Gamma_{6u}$ | 2 | 2194 | 2863 | 2964 |
|  | $\Gamma_{7u}$ | 2 | 2963 | 4412 | 3893 |
|  | $\Gamma_{8g}$ | 4 | 32267 | 34960 | 33428 |
| $5d^1$ | $\Gamma_{8g}$ | 4 | 52857 | 51091 | 52156 |
|  | $\Gamma_{7g}$ | 2 | 54395 | 52543 | 53476 |

[a] Experiment from Ref. [40].

[b] DV-Xα calculations by S. Watanabe from Ref. [17].

[c] DV-Xα calculations on $(CeF_8)^{5-}$ embedded in $CaF_2$ microcrystal in this work.

Table 6. Crystal-feild parameters and spin-orbit interaction parameters of $Ce^{3+}$ in $CaF_2$. Units are $cm^{-1}$.

|  | Parameters | Experiment[a] | Calculation[b] |
|---|---|---|---|
| 4f | $B_0^4(ff)$ | -1900 | -2041 |
|  | $B_0^6(ff)$ | 500 | 888 |
|  | $\zeta_{4f}$ | 615 | 838 |
| 5d | $B_0^4(dd)$ | -44016 | -40012 |
|  | $\zeta_{5d}$ | 1082 | 926 |

[a] Fitted parameters from Ref. [40].

[b] Calculated parameters in this work.

Table 7. 4f and 5d energy levels of $Ce^{3+}$ in $KY_3F_{10}$. Units are $cm^{-1}$. The calculated energies for the 5d states have been shifted by a constant amount so that the average matches with each other. Before this correction the calculated averages were 49038 $cm^{-1}$ and 27794 $cm^{-1}$.

|        | Calculation[a] | Calculation[b] |
|--------|----------------|----------------|
| $4f^1$ | 0              | 0              |
|        | 718            | 324            |
|        | 1579           | 518            |
|        | 2750           | 2590           |
|        | 3339           | 2860           |
|        | 3992           | 2866           |
|        | 4484           | 3371           |
| $5d^1$ | 38553          | 34367          |
|        | 46215          | 43567          |
|        | 52184          | 52994          |
|        | 53232          | 56644          |
|        | 55007          | 57619          |

[a] DV-Xα calculations from Chapter 5 of Ref. [17].

[b] DV-Xα calculations on $(CeF_8)^{5-}$ embedded in $KY_3F_{10}$ microcrystal in this work.

Table 8. Crystal-field parameters and spin-orbit interaction parameters of $Pr^{3+}$, $Nd^{3+}$ and $Ce^{3+}$ in $KY_3F_{10}$. Units are $cm^{-1}$.

| | Parameters | $Pr^{3+}$[a] | $Nd^{3+}$[b] | Calculation[c] |
|---|---|---|---|---|
| 4f | $B_0^2(ff)$ | -664 | -670 | -455 |
| | $B_0^4(ff)$ | -1543 | -1484 | -2084 |
| | $B_4^4(ff)$ | 343 | 569 | 237 |
| | $B_0^6(ff)$ | 891 | 698 | 929 |
| | $B_4^6(ff)$ | -30 | 9 | -40 |
| | $\zeta_{4f}$ | 745 | 881 | 738 |
| 5d | $B_0^2(dd)$ | | | -5192 |
| | $B_0^4(dd)$ | | | -46002 |
| | $B_4^4(dd)$ | | | -11702 |
| | $\zeta_{5d}$ | | | 832 |

[a] Fitted parameters from Ref. [54].

[b] Fitted parameters from Ref. [55].

[c] Calculated parameters in this work.

Table 9. 4f and 5d energy levels of $Ce^{3+}$ in YAG. Units are cm$^{-1}$. The calculated energies for the 5d states have been shifted by a constant amount so that the average matches the experimental average of 36570 cm$^{-1}$. Before this correction the calculated averages were 33622 cm$^{-1}$ and 24429 cm$^{-1}$ respectively.

|  | Experiment[a] | Calculation[b] | Calculation[c] |
|---|---|---|---|
| 4f$^1$ |  | 0 | 0 |
|  |  | 340 | 725 |
|  |  | 800 | 1240 |
|  |  | 2370 | 2627 |
|  |  | 2550 | 3139 |
|  |  | 2790 | 3665 |
|  |  | 4260 | 4498 |
| 5d$^1$ | 21858 | 18510 | 23497 |
|  | 29438 | 26130 | 28823 |
|  | 38314 | 42540 | 40625 |
|  | 44366 | 47010 | 42467 |
|  | 48876 | 48660 | 47439 |

[a] Experiment from Ref. [56].

[b] Calculation from Ref. [57].

[c] DV-Xα calculations on $(CeO_8)^{13-}$ embedded in YAG microcrystal in this work.

Table 10. Crystal parameters and spin-orbit interaction parameters of $Ce^{3+}$ in YAG. Units are cm$^{-1}$.

|  | Parameters | $Ce^{3+}$[a] | $Ce^{3+}$[b] | Calculation[c] |
|---|---|---|---|---|
| 4f | $B_0^2$(ff) | -465 | -380 | -275 |
|  | $B_2^2$(ff) | 96 | 261 | 449 |
|  | $B_0^4$(ff) | -3739 | -3008 | -2154 |
|  | $B_2^4$(ff) | 380 | 573 | -74 |
|  | $B_4^4$(ff) | 1602 | 1105 | 1038 |
|  | $B_0^6$(ff) | 901 | 1227 | 842 |
|  | $B_2^6$(ff) | -307 | -397 | 79 |
|  | $B_4^6$(ff) | 2136 | 1799 | 1613 |
|  | $B_6^6$(ff) | -246 | -3 | -740 |
|  | $\zeta_{4f}$ | 647 |  | 752 |
| 5d | $B_0^2$(dd) | -6099 |  | -736 |
|  | $B_2^2$(dd) | 1259 |  | 4885 |
|  | $B_0^4$(dd) | -50042 |  | -48389 |
|  | $B_2^4$(dd) | 5374 |  | 4249 |
|  | $B_4^4$(dd) | 19626 |  | 18077 |
|  | $\zeta_{5d}$ | 991 |  | 770 |

[a] Fitted parameters from Ref. [56]. Signs have been transformed to match our axis choice by rotating π/2 about z axis.

[b] Parameters estimated by C. A. Morrison, see Ref. [58].

[c] Calculated parameters in this work.

Figure 1. 5d components of calculated energy levels in a $(CeY_4Li_8F_{12})^{11+}$ cluster.

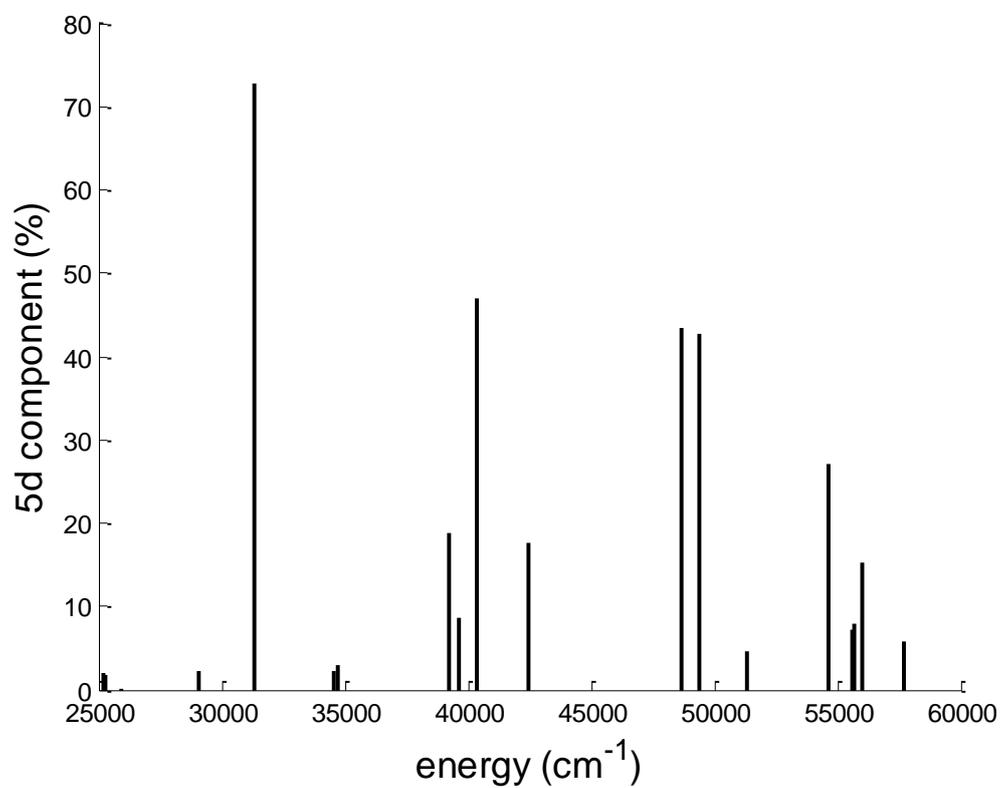